\documentclass[aps,prb,reprint]{revtex4-1}
\usepackage{amsmath,amssymb,eqnarray}
\usepackage{graphicx}
\usepackage{dcolumn}% Align table columns on decimal point
\usepackage{bm}% bold math
\usepackage{epsfig}
\usepackage{subfigure}
\usepackage[titletoc]{appendix}
\begin{document}
\title{Formation of localized magnetic states in a large-spin Fermi system}
\author{Bin-Zhou Mi$^{1,2}$, Qiang Gu$^{1}$}
\email[Corresponding author: ]{qgu@ustb.edu.cn}
\affiliation{$^1$Department of Physics and Beijing Key Laboratory for Magneto-Photoelectrical Composite and Interface Science, University of Science and Technology Beijing, Beijing 100083, China
\\
  $^2$College of Science, North China Institute of Science and Technology, Beijing 101601, China}
\date{\today}

\begin{abstract}
We extend the Anderson impurity model to a large-spin Fermi system with spin $f$=3/2, stimulated by the realization of large-spin ultracold Fermi atoms. The condition required for the spontaneous formation of local magnetic moments is examined and the ground state mean-field magnetic phase diagram is explored carefully. We find that the spin-3/2 Fermi system involves magnetic phase regions I, II, and III that correspond to one, two, and three particle/hole occupation, respectively. In addition, it is observed that all the three magnetic phases have four-fold degenerate ground states. Finally, the phase transition between the three phases seems to be of the first-order.
\end{abstract}
\maketitle

\section*{I. INTRODCTION}
Whether magnetic impurity atoms embedded in nonmagnetic metals can lead to localized magnetic states remains a challenging topic in condensed matter physics. Typically, the Anderson impurity model (AIM)\cite{1,2} is used to describe such magnetic impurities. Anderson obtained the phase boundary between the magnetic and nonmagnetic states using the variational principle\cite{1}. The local magnetic moment of the impurity atom depends sensitively on the on-site Coulomb repulsion interaction $U$, impurity atomic energy level $E_d$, and energy width 2$\Gamma$ of the localized state (also known as the virtual bound-state). Once the local magnetic moment is formed, the interaction between the electronic states of localized magnetic impurity and those of the itinerant electrons in the host metal leads to fascinating phenomena, such as Friedel oscillations\cite{3}, the Kondo effect\cite{4}, and Fano resonance\cite{5}. Actually, the physics of magnetic impurities in a metallic host has been studied intensively\cite{6,7,8,9,10,11,12,13,14,15,16,17,18,19,20} for more than fifty years and already revealed very well.

The interest in the AIM remains strong. Recently, AIM studies have been extended to finite systems\cite{21,22}, and these studies have demonstrated that the impurity spin magnetic moment increases monotonically as the host energy gap increases. Therefore, the impurity magnetic moment can be tuned by varying the size of the metal clusters and thereby adjusting their energy gaps. The AIM has also been extended to Dirac systems\cite{23,24}. A special feature of Dirac systems is that the energy band structure demonstrates a linear dispersion relation at the Dirac points, which differs from the parabolic band structure of an ordinary metal. Due to the anomalous broadening of adatom local electronic states in graphene, the local magnetic moment is formed more easily in graphene than in metals\cite{23}. Moreover, the study of three dimensional Dirac solid suggests that spin-orbit coupling facilitates the formation of localized magnetic states \cite{24}. In addition, the AIM has recently been extended to two-dimensional semiconductors with Dirac-like dispersion, and this system facilitates the formation of local magnetic moments\cite{25}.

In recent years, large-spin ultracold Fermi atoms have been realized experimentally, e.g., $^{9}$Be, $^{173}$Yb, $^{40}$K, and $^{161}$Dy, that may carry spin $f$ = 3/2, 5/2, 9/2, and 21/2 (all larger than 1/2), respectively. Quantum impurities and Kondo physics have been studied experimentally in large-spin Fermi atomic gases\cite{26,27}. Riegger et al. studied spin exchange between mobile and localized particles and simulated the coupling terms in the well-known Kondo Hamiltonian\cite{26}. In addition, Kuzmenko et al. studied the multipolar Kondo effect in an ultracold Fermi gas of $^{173}$Yb atoms and demonstrated that multipolar interactions enhance the Kondo temperature\cite{27}. In the above studies, impurity atoms were trapped in a specially designed optical potential to act as artificially controlled localized magnetic moments. However, to the best of our knowledge, the mechanism for the spontaneous formation of localized magnetic states has not been studied directly.

Therefore, this study attempts to extend the AIM to a large-spin Fermi system (LSFS) with spin $f$=3/2. We focus on the conditions required to form local magnetic moment in the LSFS which contains more spin components and thus more complex interactions between atoms\cite{28,29,30,31,32,33,34,35,36}. The LSFS is expected to present new phenomena and deepen the understanding of the physics of the AIM. The remainder of this paper is organized as follows. In Sec. II, we present the AIM Hamiltonian of the LSFS, and briefly outline formulas derived using the double-time Green's function method (DTGF) with mean-field approximation. In Sec. III, the local density of states (LDOS), ground state single-particle energy, mean-field magnetic phase diagram, and local magnetic moment are calculated and discussed carefully. At last, a brief conclusion is summarized in Sec. IV, which also includes an outlook on promising future work.

\section*{II. MODEL AND FORMULAS}
Analogous to the AIM Hamiltonian of electronic systems with spin $S$=1/2, the Anderson Hamiltonian of the LSFS with hyperfine spin $f$=3/2 can be expressed as follows\cite{1}:
\begin{align}
H&=\sum_{\textbf{k},\sigma}E_{\textbf{k}}n_{{\textbf{k}}\sigma}+\sum_{\sigma}E_dn_{d\sigma}
+\frac{U}{2}\sum^{\sigma\not=\sigma^\prime}_{\sigma,\sigma^\prime}n_{d\sigma}n_{d\sigma^\prime}\nonumber\\
&+\sum_{{\textbf{k}},\sigma}V\left(C_{{\textbf{k}}\sigma}^{\dagger}d_{\sigma}+d_{\sigma}^{\dagger}C_{{\textbf{k}}\sigma}\right)\nonumber\\
&+\frac{A_{mix}}{2}\sum^{\sigma\not=\sigma^\prime}_{\sigma,\sigma^\prime}
d_{\bar{{\sigma}}^\prime}^{\dagger}d_{\sigma^\prime}^{\dagger}d_{\bar{{\sigma}}}d_{\sigma}.
\end{align}
Here $\sigma$=$+$3/2, $+$1/2, $-$1/2, and $-$3/2 are the four internal components of hyperfine spin $f$=3/2. The first term represents the non-interacting host (free atom Fermi sea) with dispersion $E_{\textbf{k}}$, and $n_{\textbf{k}\sigma}=C_{\textbf{k}\sigma}^{\dagger}C_{\textbf{k}\sigma}$ is the number operator with wave vector $\textbf{k}$ and spin-$\sigma$. The second and third terms describe the impurity atoms (or $d$ atoms) with on-site energy $E_d$ and on-site Coulomb interaction $U$. $E_d$ defines the relative position of the $d$ atomic state relative to the Fermi energy, and $n_{d\sigma}=d_{\sigma}^{\dagger}d_{\sigma}$ is the number operator of the $d$ atomic state with spin-$\sigma$. The $d$ atomic state is used to maintain consistency with the $d$ electronic state in condensed matter systems. The fourth term describes one-atom with spin-$\sigma$ hybridization between the free atomic and $d$ atomic states. $V$ is the hybridization strength, which is assumed to be independent of $\textbf{k}$.

The last term in Eq. (1) represents the spin-exchange interactions of $d$ atoms that mix different spin components\cite{30, 34, 35}, and $A_{mix}$ is the spin-mixing strength. For example, two atoms with hyperfine spins of $-$1/2 and $+$1/2 may change into two atoms with hyperfine spins of $-$3/2 and $+$3/2, and vice versa. In cold atomic systems, spin-mixing strength can be adjusted experimentally. For simplicity, we consider the $A_{mix}$=0 case in the present paper.

The DTGF method is a powerful tool to study quantum many body systems. In order to study the occurrence of localized magnetic moments in the model given by Eq. (1), the following retarded Green's function (GF) defined by operators $A$ and $B$ at times $t$ and $t^\prime$, respectively, is considered:
\begin{align}
\langle\langle A(t)|B(t^\prime)\rangle\rangle^R&=-i\theta(t-t^\prime)\langle[A(t),B(t^\prime)]_+\rangle,
\end{align}
where $\theta(x)$ is the unit step function. We denote $[A,B]_+=AB+BA$ and refer to the GF as the Fermionic GF. Then, the derivative of Eq. (2) with respect to $t$ is calculated. After performing time Fourier transformation, the equation of motion for the GF is obtained as follows:
\begin{align}
\omega\langle\langle A|B\rangle\rangle_\omega&=\langle[A,B]_+\rangle+\langle\langle[A,H]|B\rangle\rangle_\omega.
\end{align}

By applying Eq. (3) to the $d$ atomic GF $\langle\langle d_{\sigma}|d_{\sigma}^{\dagger}\rangle\rangle_\omega$, the mean-field approximation\cite{1} is used to obtain the following expression:
\begin{align}
\left(\omega-E_d-U\sum^{\sigma^\prime\not=\sigma}_{\sigma^\prime}\langle n_{d\sigma^\prime}\rangle\right)\langle\langle
d_{\sigma}|d_{\sigma}^{\dagger}\rangle\rangle_\omega=1+\sum_{\textbf{k}}V\langle\langle C_{\textbf{k}\sigma}|d_{\sigma}^{\dagger}\rangle\rangle_\omega.
\end{align}
The mixed GF $\langle\langle C_{\textbf{k}\sigma}|d_{\sigma}^{\dagger}\rangle\rangle_\omega$ appears in Eq. (4). Similarly, by applying Eq. (3) to the GF $\langle\langle C_{\textbf{k}\sigma}|d_{\sigma}^{\dagger}\rangle\rangle_\omega$, we obtain
\begin{align}
\left(\omega-E_\textbf{k}\right)\langle\langle C_{\textbf{k}\sigma}|d_{\sigma}^{\dagger}\rangle\rangle_\omega&=V\langle\langle
d_{\sigma}|d_{\sigma}^{\dagger}\rangle\rangle_\omega.
\end{align}
By inserting Eq. (5) into Eq. (4), we obtain the solutions of the $d$ atomic GF as follows:
\begin{align}
\langle\langle d_{\sigma}|d_{\sigma}^{\dagger}\rangle\rangle_{\omega+i0^+}&=\frac{1}{\left(\omega-E_d-U\sum^{\sigma^\prime\not=\sigma}_{\sigma^\prime}\langle n_{d\sigma^\prime}\rangle\right)+i\Gamma},
\end{align}
and
\begin{align}
\langle\langle d_{\sigma}|d_{\sigma}^{\dagger}\rangle\rangle_{\omega-i0^+}&=\frac{1}{\left(\omega-E_d-U\sum^{\sigma^\prime\not=\sigma}_{\sigma^\prime}\langle n_{d\sigma^\prime}\rangle\right)-i\Gamma},
\end{align}
where $\Gamma=\pi V^2\rho_{F}^{(0)}$, and $\rho_{F}^{(0)}$ is the density of states of free atoms at Fermi energy $E_F$. We take $\rho_{F}^{(0)}$ as a constant.

The ensemble average  $\langle n_{d\sigma}\rangle=\langle d_{\sigma}^{\dagger}d_{\sigma}\rangle$ is calculated self-consistently from $d$ atomic GFs through the spectral theorem:
\begin{align}
\langle BA\rangle&=\frac{i}{2\pi}\int_{-\infty}^{\infty}\left\{\langle\langle A|B\rangle\rangle_{\omega+i0^+}-\langle\langle A|B\rangle\rangle_{\omega-i0^+}\right\}f(\omega)\, d\omega,
\end{align}
where $f(\omega)$ is the Fermi distribution function. At zero temperature, the occupation is expressed as follows:
\begin{align}
\langle n_{d\sigma}\rangle&=\int_{-\infty}^{E_F} \rho_{d\sigma}(\omega)\, d\omega,
\end{align}
and the LDOS of $d$ atoms with spin-$\sigma$ is given as:
\begin{align}
\rho_{d\sigma}(\omega)&=\frac{1}{\pi}\frac{\Gamma}{\left(\omega-E_d-U\sum^{\sigma^\prime\not=\sigma}_{\sigma^\prime}\langle n_{d\sigma^\prime}\rangle\right)^2+\Gamma^2}.
\end{align}
The peak value of the LDOS of $d$ atoms is $1/\pi\Gamma$, and the energy width of a virtual bound-state is 2$\Gamma$. By inserting Eq. (10) into Eq. (9), we obtain
\begin{align}
\langle n_{d\sigma}\rangle&=\frac{1}{\pi}\cot^{-1}\left(\frac{E_d+U\sum^{\sigma^\prime\not=\sigma}_{\sigma^\prime}\langle n_{d\sigma^\prime}\rangle-E_F}{\Gamma}\right),
\end{align}
or
\begin{align}
\sum^{\sigma^\prime\not=\sigma}_{\sigma^\prime}\langle n_{d\sigma^\prime}\rangle&=\frac{\Gamma}{U}\cot[\pi\langle n_{d\sigma}\rangle]+\frac{E_F-E_d}{U}.
\end{align}

Eq. (12) is a self-consistent equation set comprising four equations. The local magnetic moment of $d$ atoms is defined as
\begin{align}
m&=(\langle n_{d\frac{1}{2}}\rangle-\langle n_{d-\frac{1}{2}}\rangle)+3(\langle n_{d\frac{3}{2}}\rangle-\langle n_{d-\frac{3}{2}}\rangle),
\end{align}
and the total occupation number of $d$ atoms is
\begin{align}
n_{td}&=\langle n_{d\frac{1}{2}}\rangle+\langle n_{d-\frac{1}{2}}\rangle+\langle n_{d\frac{3}{2}}\rangle+\langle n_{d-\frac{3}{2}}\rangle.
\end{align}

By tuning the parameters, e.g., $E_F$, $E_d$, $\Gamma$ and $U$, a set of trivial solutions (non-magnetic solutions) can always be obtained by Eq. (12), denoted as $\langle n_{d\frac{1}{2}}\rangle=\langle n_{d-\frac{1}{2}}\rangle=\langle n_{d\frac{3}{2}}\rangle=\langle n_{d-\frac{3}{2}}\rangle=n_0$. However, nontrivial solutions (magnetic solutions) can only be obtained in the magnetic regions\cite{1}. The condition for the formation of localized magnetic states is that the occupation numbers of the four internal spin components of $d$ atoms are not all equal. Differing from the spin-1/2 AIM, the spin structure of the spin-3/2 impurity leads to a large number of possible magnetic ground states, and the phase boundary between the magnetic and nonmagnetic states of spin-3/2 AIM cannot be obtained directly via the variational principle\cite{1}. The magnetic solutions are obtained by numerical calculations in Section III. The numerical results demonstrate that there are several types of magnetic solutions for the self-consistent equations.

Which type of magnetic solution is the most stable? Herein, two schemes are proposed to evaluate the stability of the magnetic solutions. In the first scheme, the phase boundary between the magnetic and nonmagnetic states can be determined effectively by analyzing the variation of the LDOS of the $d$ atoms. In the second scheme, the stability of different magnetic solutions is analyzed from an energy perspective. Free atoms form a Fermi sea, and the $d$ atoms are akin to a drop of water in the greater Fermi sea. Therefore, the ground state energy $E_0$ of $d$ atoms is calculated to determine the most feasible magnetic solution. In fact, for a set of parameters in the magnetic regions, the total numbers of $d$ atoms for different types of magnetic solutions are unequal, but there is a fine distinction between them. Thus, the single-particle energy $E_{0S}$ of $d$ atoms is used to evaluate the stability of the magnetic solution.

We analyzed the reliability of the method to determine the phase boundary. The main criterion is based on the calculation of the ground state single-particle energy. We applied this criterion to check the phase boundary of the spin-1/2 system. Calculation shows that the single-particle energy of the magnetic solution is always lower than that of the nonmagnetic solution in the magnetic region. The obtained phase boundary is completely consistent with Anderson$^{,}$s original paper\cite{1}. Therefore, we believe that applying this criterion to the spin-3/2 Fermi system is reasonable. This criterion can not only determine the outer boundary between the magnetic and nonmagnetic states, but can also evaluate the inner boundary between the three magnetic phases.

From Eq. (10), the LDOS of $d$ atoms is a lorentzian-type curve centered at $E_d+U\sum^{\sigma^\prime\not=\sigma}_{\sigma^\prime}\langle n_{d\sigma^\prime}\rangle$. In other words, virtual bound-states are formed at the following energies:
\begin{align}
\omega_{d\sigma}&=E_d+U\sum^{\sigma^\prime\not=\sigma}_{\sigma^\prime}\langle n_{d\sigma^\prime}\rangle.
\end{align}
Thus, the ground state energy of $d$ atoms can be defined as follows:
\begin{align}
E_0&=\sum_{\sigma}\omega_{d\sigma}\langle n_{d\sigma}\rangle.
\end{align}
Finally, the single-particle energy of $d$ atoms is expressed as follows:
\begin{align}
E_{0S}&=\frac{E_0}{n_{td}}.
\end{align}

The Fermi energy $E_F$ is set to zero ($E_F$=0) in the following numerical calculations. $U$=1 is fixed as the unit of the system, and the $E_d$ and $\Gamma$ values are varied. All parameters are considered as dimensionless quantities.

\section*{III. NUMERICAL RESULTS AND DISCUSSION}
\subsection{Mean-field magnetic phase diagram for particles}
This study primarily aims to explore the ground state magnetic phase diagram. The basic idea is to obtain all possible magnetic solutions by solving nonlinear Eq. (12) numerically and then classifying the obtained magnetic solutions. Several types of magnetic solutions exist for a given set of parameters. The most stable magnetic solution is obtained by comparing the LDOS and the single-particle energy of $d$ atoms for different types of magnetic solutions.

First, the LDOS of $d$ atoms is calculated as a function of energy $\omega$. Prior to selecting a set of specific parameter values, symmetry analysis of the solutions is performed. From the LDOS function of $d$ atoms in Eq. (10) and the ground state energy function in Eq. (16), it is evident that the systems may have a degenerate ground state because the ground state energy does not change as the order of the hyperfine spin-$\sigma$ subscripts \{$+$3/2, $+$1/2, $-$1/2, $-$3/2\} is exchanged arbitrarily for a certain type of magnetic solution.

Consider the set of parameter values $(E_F-E_d)/U$=0.5 and $\pi\Gamma/U$=0.2. Note that the parameters are reduced by $U$. The magnetic solution of Eq. (12) comprises three categories: type 1 $\left(\{\langle n_{d\sigma}\rangle\}=\{0.9478, 0.0384, 0.0384, 0.0384\}\right)$, type 2 $\left(\{\langle n_{d\sigma}\rangle\}=\{0.4115, 0.4115, 0.0533, 0.0533\}\right)$, and type 3 $\left(\{\langle n_{d\sigma}\rangle\}=\{0.2385, 0.2385, 0.2385, 0.0914\}\right)$. Similar to the above analysis, each type of magnetic solution demonstrates rotational symmetry.

Figures 1(a)-(c) plot the LDOS of $d$ atoms versus energy $\omega$ for the type 1, 2, and 3 solutions, respectively. In this case, the peak values of all LDOS are $1/\pi\Gamma$=5; the energy widths of all possible virtual bound-states are $2\Gamma$. Figure 1(a) indicates that one majority state is located below the Fermi energy, while the other three minority states are above the Fermi energy. In addition, Fig. 1(b) shows that two majority states are above and very close to the Fermi energy, and the other two minority states are far above the Fermi energy. Furthermore, Fig. 1(c) demonstrates that three majority states and one minority state are located above the Fermi energy. Both the majority and minority spin states clearly feature a Lorentzian-type state density curve, which is similar to the spin-1/2 Anderson impurity system\cite{1}. In particular, as can be seen in Figs. 1(a)-(c), only the majority state in Fig. 1(a) is below the Fermi energy, which means that the type 1 solution has the lowest energy; therefore, it is the most stable. This is also confirmed by calculating the single-particle energy $E_{0S}$ for the three types of magnetic solutions.

Figure 2 plots the LDOS of $d$ atoms versus energy $\omega$ for several values of $\pi\Gamma/U$ for the three types of magnetic solutions at $(E_F-E_d)/U$=0.5. By comparing Figs. 2(a), 2(b), and 2(c), it is evident that the type 1 solution is always the most stable. Figure 2(a) indicates that as hybridization strength increases, the energy width of the virtual bound-state and the height of the peak of the LDOS of $d$ atoms increase and decrease, respectively. The peak position of the LDOS function shifts toward the Fermi surface. This behavior can be understood as follows. From the energy-time uncertainty relation, the energy width of the virtual bound-state is inversely proportional to the life time of $d$ atoms. Therefore, from Fig. 2(a), it can be concluded that the $d$ atoms becomes more delocalized as $\pi\Gamma$ increases. Thus, the localized magnetic states decrease. Evidently, when $\pi\Gamma$ is increased to a critical value, the distance between the centers of the LDOS of the majority and the minority states is zero, the solution of Eq. (12) becomes trivial and the localized magnetic state disappears; this corresponds to the phase boundary between the magnetic and nonmagnetic states. In contrast, the effect of $U$ should be the opposite.

\begin{figure}
    \centering
    \includegraphics[width=0.4\textwidth]{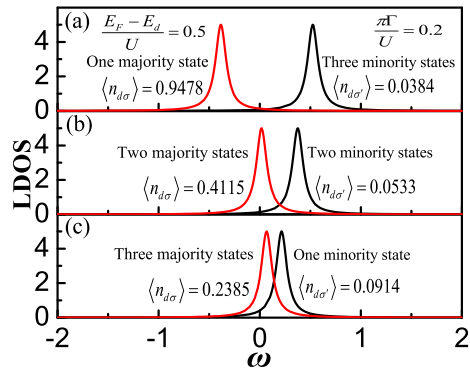}
    \caption{(Color online) LDOS of $d$ atoms versus energy $\omega$ for parameters $(E_F-E_d)/U=0.5$ and $\pi\Gamma/U=0.2$. (a), (b), and (c) denote type 1, 2, and 3 magnetic solutions, respectively.}
\end{figure}
\begin{figure}
    \centering
    \includegraphics[width=0.4\textwidth]{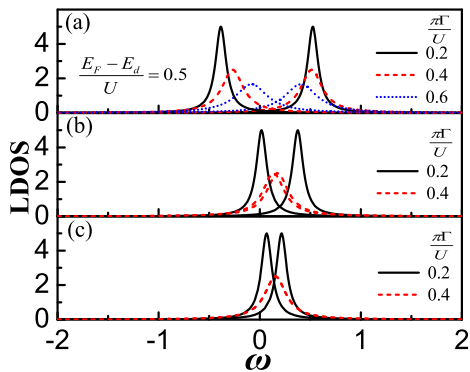}
    \caption{(Color online) LDOS of $d$ atoms versus energy $\omega$ for several values of $\pi\Gamma/U$ at $(E_F-E_d)/U=0.5$. (a), (b), and (c) denote type 1, 2, and 3 magnetic solutions, respectively. Here, the type 1 solution is the most stable.}
\end{figure}

Now let's take another set of parameters $(E_F-E_d)/U$=2.5 and $\pi\Gamma/U$=0.2 for calculation. The corresponding magnetic solution of Eq. (12) also contains three categories, i.e., type 1 $\left(\{\langle n_{d\sigma}\rangle\}=\{0.9086, 0.7615, 0.7615, 0.7615\}\right)$, type 2 $\left(\{\langle n_{d\sigma}\rangle\}=\{0.9467, 0.9467, 0.5885, 0.5885\}\right)$, and type 3 $\left(\{\langle n_{d\sigma}\rangle\}=\{0.9616, 0.9616, 0.9616, 0.0522\}\right)$. Each type of solution also demonstrates rotational symmetry.

Figures 3(a)-(c) plot the LDOS of $d$ atoms versus energy $\omega$ for the type 1, 2, and 3 solutions, respectively. Figure 3(a) shows that one majority state and the other three minority states are located below the Fermi energy. Figure 3(b) exhibits that two majority states are located below the Fermi energy and the other two minority states are below and very close to the Fermi energy. Figure 3(c) indicates that three majority states are located far below the Fermi energy and one minority state is above the Fermi energy. In addition, the distance between the centers of the LDOS of the majority and minority states in Fig. 3(c) is the greatest. Moreover, the corresponding spin state of Fig. 3(c) has the lowest single-particle energy $E_{0S}$; thus, the type 3 magnetic solution is the most stable. Figure 4 plots the LDOS of the $d$ atoms as a function of energy $\omega$ for three $\pi\Gamma/U$ values at $(E_F-E_d)/U$=2.5. In this case, the type 3  magnetic solution is always the most stable. Similar to Fig. 2(a), Fig. 4 indicates that as $\pi\Gamma$ increases, the energy width and height of the peak of the LDOS of $d$ atoms increase and decrease, respectively, and the peak position of the LDOS function moves toward the Fermi energy. The reasons for these results are the same as those described previously.

\begin{figure}
    \centering
    \includegraphics[width=0.4\textwidth]{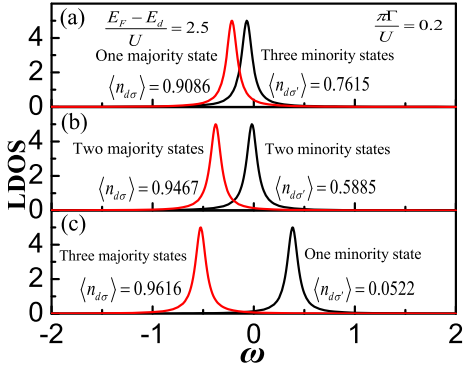}
    \caption{(Color online) LDOS of $d$ atoms versus energy $\omega$ for parameters $(E_F-E_d)/U=2.5$ and $\pi\Gamma/U=0.2$. (a), (b), and (c) denote type 1, 2, and 3 magnetic solutions, respectively.}
\end{figure}
\begin{figure}
    \centering
    \includegraphics[width=0.4\textwidth]{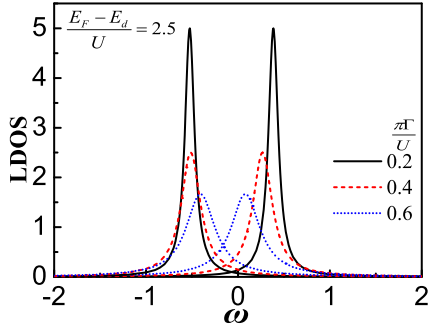}
    \caption{(Color online) LDOS of $d$ atoms versus energy $\omega$ for three $\pi\Gamma/U$ values with $(E_F-E_d)/U=2.5$. Here, the type 3 solution is the most stable.}
\end{figure}

In the previous calculations, parameter values $(E_F-E_d)/U$=0.5 and 2.5 were used. In the following, we consider a symmetric case where $(E_F-E_d)/U$=1.5. For parameters $(E_F-E_d)/U$=1.5 and $\pi\Gamma/U$=0.2, the magnetic solution of Eq. (12) contains four categories, i.e., type 1 $\left(\{\langle n_{d\sigma}\rangle\}=\{0.9678, 0.2907, 0.2907, 0.2907\}\right)$, type 2 $\left(\{\langle n_{d\sigma}\rangle\}=\{0.9558, 0.9558, 0.0442, 0.0442\}\right)$, type 3 $\left(\{\langle n_{d\sigma}\rangle\}=\{0.7093, 0.7093, 0.7093, 0.0322\}\right)$, and type 4 $\left(\{\langle n_{d\sigma}\rangle\}=\{0.9558, 0.5, 0.5, 0.0442\}\right)$.

Figures 5(a)-(d) plot the LDOS of $d$ atoms versus energy $\omega$ for these type 1, 2, 3, and 4 solutions, respectively. Figure 5(a) shows that a single majority state is located far below the Fermi energy and that the other three minority states are above and very close to the Fermi energy. Bilateral symmetry, Fig. 5(c) indicates that three majority states are below and very close to the Fermi energy and that the other minority state is above and far higher than the Fermi energy. Furthermore, the distances between the centers of the LDOS of the majority and minority states of Figs. 5(a) and 5(c) are equal. Figure 5(b) demonstrates that the two majority states are located below the Fermi energy and the other two minority states are above the Fermi energy. Furthermore, the centers of the LDOS of the majority and minority states in Fig. 5(b) are symmetrical relative to the Fermi energy. Figure 5(d) displays that one majority state is below the Fermi energy, the other minority state is above the Fermi energy, and two other median states are just at the Fermi energy. Importantly, the distances between the centers of the LDOS of the majority and minority states in Figs. 5(b) and 5(d) are also equal, which are greater than the corresponding distances in Figs. 5(a) and 5(c). Finally, from the energy perspective, the single-particle energy $E_{0S}$ of the states in Fig. 5(b) is the lowest; thus, the type 2 solution is the most stable.

Figure 6 plots the LDOS of $d$ atoms as a function of energy $\omega$ for several $\pi\Gamma/U$  values at $(E_F-E_d)/U$=1.5. In this case, the type 2 solution is always the most stable. Similar to Figs. 2(a) and 4, Fig. 6 indicates that the peak position of the LDOS function moves toward the Fermi energy as $\pi\Gamma$ increases. When $\pi\Gamma$ increases to a critical value, the distance between the centers of the LDOS of the majority and minority states is zero, the type 2 solution becomes trivial and the local magnetic moment disappears, which corresponds to the phase boundary between the magnetic and nonmagnetic states.

\begin{figure}
    \centering
    \includegraphics[width=0.45\textwidth]{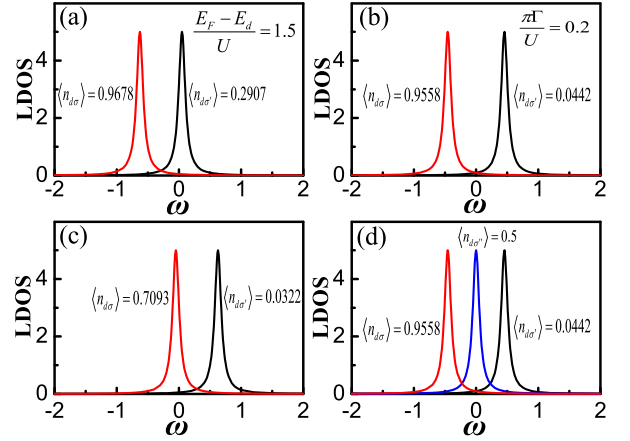}
    \caption{(Color online) LDOS of $d$ atoms versus energy $\omega$ for parameters $(E_F-E_d)/U=1.5$ and $\pi\Gamma/U=0.2$. (a), (b), (c), and (d) denote type 1, 2, 3, and 4 magnetic solutions, respectively.}
\end{figure}
\begin{figure}
    \centering
    \includegraphics[width=0.4\textwidth]{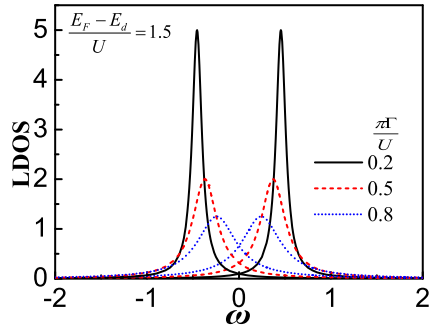}
    \caption{(Color online) LDOS of $d$ atoms versus energy $\omega$ for three $\pi\Gamma/U$  values at $(E_F-E_d)/U=1.5$. In this case, the type 2 magnetic solution is the most stable.}
\end{figure}

Based on the above calculations, parameter $(E_F-E_d)/U$ is fixed, the $\pi\Gamma/U$ values change, and the most stable magnetic solution and the boundary between the magnetic and nonmagnetic states are obtained by analyzing the LDOS and single-particle energy. In the following, the parameter $\pi\Gamma/U$ is fixed and the $(E_F-E_d)/U$ values are varied to calculate the transition boundaries between different magnetic solutions. Figure 7 plots the single-particle energy $E_{0S}$ versus $(E_F-E_d)/U$ for different types of magnetic solutions at $\pi\Gamma/U$=0.2. Figure 7 shows that the type 4 solution only exists within the narrow range of parameter $(E_F-E_d)/U$, which is always unfavorable in terms of energy. As shown in Fig. 7, there are two transition points, one is at $(E_F-E_d)/U$=1.4012539, where the type 1 solution is transformed into the type 2 solution, and the other is at $(E_F-E_d)/U$=2.3153605, where the type 2 solution turns into the type 3 solution.

\begin{figure}
    \centering
    \includegraphics[width=0.4\textwidth]{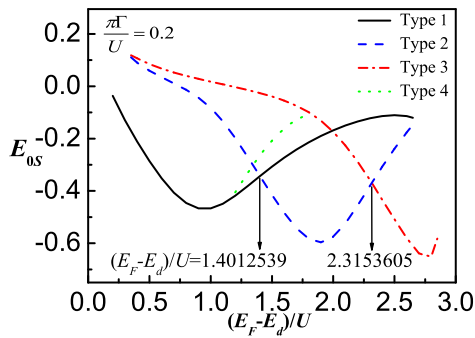}
    \caption{(Color online) Single-particle energy $E_{0S}$ of $d$ atoms versus $(E_F-E_d)/U$ for different types of magnetic solutions at $\pi\Gamma/U=0.2$.}
\end{figure}

\begin{figure}
    \centering
    \includegraphics[width=0.4\textwidth]{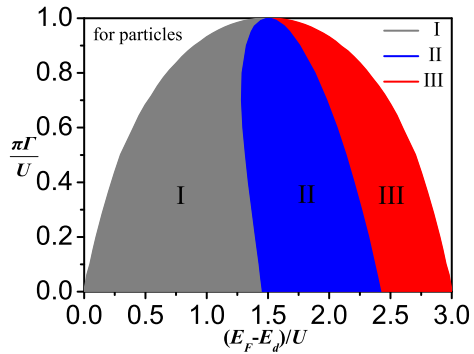}
    \caption{(Color online) Mean-field magnetic phase diagram for particles.}
\end{figure}

To understand the formation of localized magnetic states in the LSFS comprehensively, the mean-field magnetic phase diagram is calculated in Fig. 8. Here, the two reductive independent parameters $(E_F-E_d)/U$ and $\pi\Gamma/U$ are selected as the variables in the parameter space. The numerical results demonstrate that there are three magnetic phases: I, II, and III, which correspond to the type 1, 2, and 3 solutions, respectively. In the phase I region, the type 1 solution is the most stable. It can be seen from Fig. 1(a) that the type 1 solution contains one majority spin state and three minority spin states. The total occupation number $n_{td}$ is 1.063 (close to 1), which corresponds to four-fold degenerate single-occupancy states with energy $E_d$. For the phase III region, the type 3 solution is the most stable. As shown in Fig. 3(c), the type 3 solution contains three majority spin states and one minority spin state. The total occupation number $n_{td}$ is 2.937 (close to 3), which corresponds to four-fold degenerate three-occupancy states with energy $3E_d+3U$. In the phase II region, the type 2 solution is the most stable. Figure 5(b) shows that the type 2 solution contains two majority spin states and two minority spin states. Here, the total occupation number $n_{td}$ is 2, which corresponds to a double-occupancy state. There are four-fold degenerate double-occupancy states (two nonmagnetic states are not considered here) with energy $2E_d+U$. Finally, summing up the above discussions, one can see that the magnetic phases I, II, and III correspond to one, two, and three particle occupation, respectively.

Indeed, parameters $\pi\Gamma/U$ and $(E_F-E_d)/U$ in the phase diagram have physical meanings. Obviously, as $\pi\Gamma/U$ increases, the region of the localized magnetic states becomes smaller. The reasons for this can be understood as follows: the large $\pi\Gamma/U$ is equivalent to small $U$ and large hybridization $\Gamma$, that is, the increment of $\pi\Gamma/U$ weakens the correlation between the $d$ atoms, while enhances hybridization between the $d$ atoms and the Fermi sea (itinerant atom). As a result, $d$ atoms become more delocalized, thereby making the formation of localized magnetic states more difficult. In the phase diagram, parameter $(E_F-E_d)/U=0$ and 3 ($\pi\Gamma/U$=0) correspond to the empty $d$ atomic and four-particle states, respectively, and $0<(E_F-E_d)/U<3$ is the only magnetic range.

To proceed, we analyze the symmetry of the phase diagram. The phase boundary between the magnetic and nonmagnetic states is symmetric relative to $(E_F-E_d)/U=1.5$ due to the particle-hole (PH) symmetry\cite{37,38,39}, which is similar to the case of spin-1/2 AIM\cite{1}. In contrast, the phase boundaries between the three magnetic phases are asymmetric. In addition, we calculate the magnetic phase diagram in the hole representation, as shown in the Appendix. The obtained results show that the two phase-diagrams for the particle and the hole are symmetrical with $(E_F-E_d)/U=1.5$ as a line axis, which reflects, to some extent, the PH symmetry of the model.

\subsection{Occupation number and local magnetic moment of $d$ atoms}

\begin{figure}
    \centering
    \includegraphics[width=0.4\textwidth]{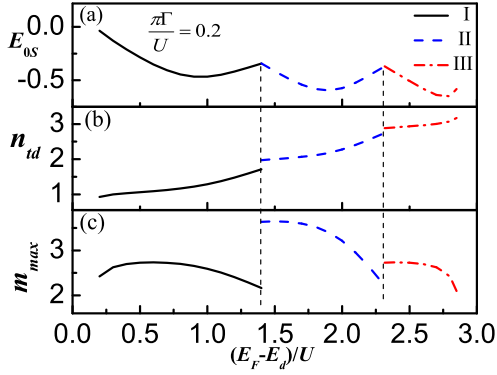}
    \caption{(Color online) (a) Single-particle energy $E_{0S}$, (b) total occupation number $n_{td}$, and (c) maximum local magnetic moment $m_{max}$ of $d$ atoms versus $(E_F-E_d)/U$ at $\pi\Gamma/U=0.2$.}
\end{figure}
\begin{figure}
    \centering
    \includegraphics[width=0.4\textwidth]{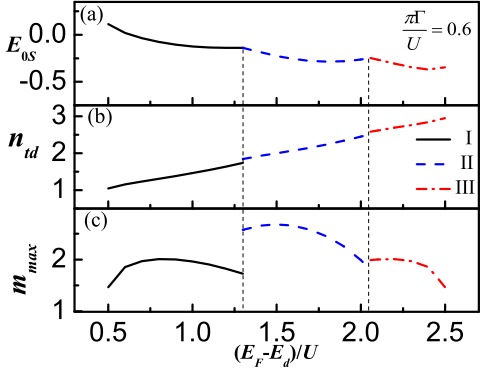}
    \caption{(Color online) (a) Single-particle energy $E_{0S}$, (b) total occupation number $n_{td}$, and (c) maximum local magnetic moment $m_{max}$ of $d$ atoms versus $(E_F-E_d)/U$ at $\pi\Gamma/U=0.6$.}
\end{figure}

In this subsection, we focus on the magnetic properties of three magnetic phases. Firstly, an inherent question is whether the transition between different magnetic phases is of continuous phase transition. Figures 9 and 10 show the variation curves of single-particle energy $E_{0S}$, total occupation number $n_{td}$, and maximum local magnetic moment $m_{max}$ of $d$ atoms with $(E_F-E_d)/U$ at $\pi\Gamma/U$=0.2 and 0.6, respectively. As can be seen from Figs. 9 and 10, a discontinuity can be observed in both the total occupation number and local magnetic moment of $d$ atoms, which means that the transition between the three phases might be a first-order phase transition. Furthermore, by comparing Figs. 9(b) and 10(b), we observe that the discontinuity gap of the total occupation number of $d$ atoms decreases with increasing $\pi\Gamma/U$. In addition, Figs. 9(c) and 10(c) indicate that the maximum $m_{max}$ occurs at $(E_F-E_d)/U$=1.5. As shown in Fig. 8, the maximum range of localized magnetic states also appears at $(E_F-E_d)/U$=1.5.

\begin{figure}
    \centering
    \includegraphics[width=0.45\textwidth]{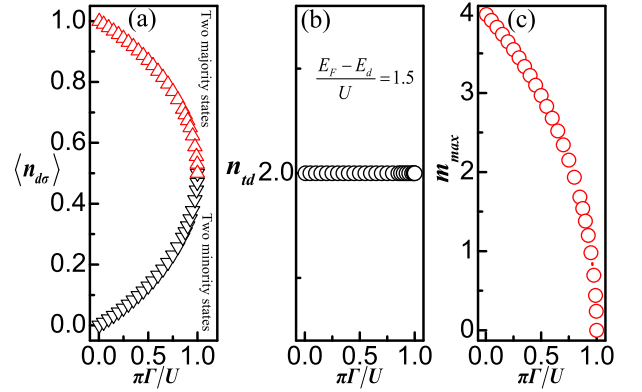}
    \caption{(Color online) Occupation number and local magnetic moment of $d$ atoms versus $\pi\Gamma/U$ at $(E_F-E_d)/U=1.5$. (a) Occupation number of different spin components of $d$ atoms (corresponding to the type 2 magnetic solution). (b) Total occupation number $n_{td}$ of $d$ atoms. (c) Maximum localized magnetic moments $m_{max}$ of $d$ atoms.}
\end{figure}
\begin{figure}
    \centering
    \includegraphics[width=0.45\textwidth]{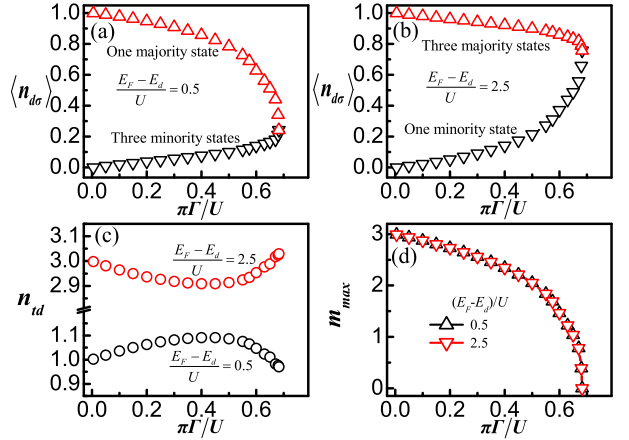}
    \caption{(Color online) Comparison of the occupation number and local magnetic moment of $d$ atoms versus $\pi\Gamma/U$ for $(E_F-E_d)/U=0.5$ and 2.5. Occupation number of different spin components of $d$ atoms for (a) $(E_F-E_d)/U=0.5$ (corresponding to the type 1 magnetic solution) and (b) $(E_F-E_d)/U=2.5$ (corresponding to the type 3 magnetic solution). (c) Total occupation numbers $n_{td}$ of $d$ atoms for $(E_F-E_d)/U=0.5$ and 2.5. (d) Maximum localized magnetic moments $m_{max}$ of $d$ atoms for $(E_F-E_d)/U=0.5$ and 2.5.}
\end{figure}

The influences of $\pi\Gamma/U$ on the occupation number of different spin components, total occupation numbers, and localized magnetic moments of $d$ atoms are calculated in Fig. 11 for $(E_F-E_d)/U$=1.5 and Fig. 12 for $(E_F-E_d)/U$=0.5 and 2.5. Figures 11(a), 12(a), and 12(b) show that for fixed $(E_F-E_d)/U$, the occupation number of the majority spin state decreases with increasing $\pi\Gamma/U$. In contrast, the occupation number of the minority spin state increases with increasing $\pi\Gamma/U$. Therefore, as shown in Figs. 11(c) and 12(d), the maximum local magnetic moment decreases with increasing $\pi\Gamma/U$ and seems to drop sharply to zero at the critical boundary point. One can take $m_{max}$ as an order parameter, a critical exponent $\beta\approx1/2$ is obtained by numerical calculation, which is the result of the mean field.

More importantly, in Fig. 12(b), we observe that under fixed $(E_F-E_d)/U$=1.5, the total occupation number of $d$ atoms $n_{td}$ is 2 (half-filled) for any $\pi\Gamma/U$ value ($0\leq\pi\Gamma/U\leq1$). This situation corresponds to the double-occupancy state of the $d$ atoms and seems to form localized magnetic states easily from magnetic phase diagram. In Fig. 12(c), for $(E_F-E_d)/U$=0.5, the total occupation number of $d$ atoms $n_{td}$ slightly increases initially and then decreases slowly with increasing $\pi\Gamma/U$; however, all changes are in the vicinity of 1, i.e., this is a single-occupancy state of the $d$ atoms. For $(E_F-E_d)/U$=2.5, the total occupation number of $d$ atoms $n_{td}$ first decreases slightly and then increases slowly with increasing $\pi\Gamma/U$; however, all changes are in the vicinity of 3, which corresponds to the three-occupancy state of the $d$ atoms. As shown in Figs. 1(a), 3(c), 8, and 12(a)-(d), the data for the $(E_F-E_d)/U$=0.5 and 2.5 cases are related via PH symmetry relative to $n_{td}=2$, i.e., the sum of the total occupancy of their $d$ atoms is 4 for arbitrary $\pi\Gamma/U$; thus, the local magnetic moment curves overlap completely.

\section*{IV. CONCLUSIONS}
In this study, based on the AIM, the formation of localized magnetic states in a spin-3/2 itinerant Fermi system is investigated by employing the DTGF method with mean-field approximation. The magnetic phase diagram in the $((E_F-E_d)/U, \pi\Gamma/U)$ plane is presented numerically. It is found that the spin-3/2 Anderson impurity system has much richer phases than that of the spin-1/2 system. There are three magnetic phases I, II, and III that correspond to one, two, and three particle occupation, respectively. In addition, each magnetic phase has a four-fold degenerate ground state without an external magnetic field. Moreover, our calculations suggest that the phase transition between the three phases might be of the first-order.

The spin-mixing term is not considered in this paper. This term may result in a nontrivial effect on the magnetic impurity formation and will be studied in future. For example, the subscript-$\sigma$ in the $d$ atomic Green's function loses rotational symmetry when spin-mixing term is taken into account, and thus may reduce or eliminate the degeneracy of the three phases.

\begin{acknowledgments}
Bin-Zhou Mi thanks Junjun Xu, Jingxiang Zhao, and Zongli Sun for their useful suggestions. This study was supported by the National Natural Science Foundation of China (Grant No. 11574028), the Higher Educational Science and Technology Research Project of Youth Fund of Hebei Province (Grant No. QN2018301), the Natural Science Foundation of Shandong Province of China (Grant No. ZR2017MEM012), and the Fundamental Research Funds for the Central Universities, North China Institute of Science and Technology (Grant No. 3142017069).
\end{acknowledgments}

\setcounter{equation}{0}
\renewcommand\theequation{A\arabic{equation}}
\begin{appendices}
\section*{APPENDIX: Mean-field magnetic phase diagram for holes}
In the Appendix, the ground state mean-field magnetic phase diagram for holes is calculated. The PH transformation can be defined as: $C_{\textbf{k}\sigma}^{\dagger}\rightarrow C_{\textbf{-k}\sigma}, C_{\textbf{k}\sigma}\rightarrow C_{\textbf{-k}\sigma}^{\dagger}, d_{\sigma}^{\dagger}\rightarrow $-$d_{\sigma}, d_{\sigma}\rightarrow $-$d_{\sigma}^{\dagger}$. In the hole representation, the Hamiltonian expressed by Eq. (1) with $A_{mix}$=0 becomes
\begin{align}
H&=\sum_{\textbf{-k},\sigma}E_{\textbf{-k}}C_{\textbf{-k}\sigma}C_{\textbf{-k}\sigma}^{\dagger}+\sum_{\sigma}E_dd_{\sigma}d_{\sigma}^{\dagger}
+\frac{U}{2}\sum^{\sigma\not=\sigma^\prime}_{\sigma,\sigma^\prime}d_{\sigma}d_{\sigma}^{\dagger}d_{\sigma^\prime}d_{\sigma^\prime}^{\dagger}\nonumber\\
&+\sum_{{\textbf{-k}},\sigma}V\left(d_{\sigma}^{\dagger}C_{{\textbf{-k}}\sigma}+C_{{\textbf{-k}}\sigma}^{\dagger}d_{\sigma}\right).
\end{align}
For the hole, the similar calculation and analysis as the particle are performed. Firstly, the LDOS of $d$ holes with spin-$\sigma$ can be given as:
\begin{align}
\rho_{d\sigma}(\omega)&=\frac{1}{\pi}\frac{\Gamma}{\left[\omega+E_d+U\sum^{\sigma^\prime\not=\sigma}_{\sigma^\prime}\left(1-\langle n_{d\sigma^\prime}\rangle\right)\right]^2+\Gamma^2}.
\end{align}
Accordingly, the occupation number $\langle n_{d\sigma}\rangle$ of $d$ holes is obtained by the following self-consistent equation set:
\begin{align}
\sum^{\sigma^\prime\not=\sigma}_{\sigma^\prime}\left(1-\langle n_{d\sigma^\prime}\rangle\right)&=\frac{\Gamma}{U}\cot[\pi\left(1-\langle n_{d\sigma}\rangle\right)]+\frac{E_F-E_d}{U}.
\end{align}
From Eq. (A2), virtual bound-states are formed at the following energies:
\begin{align}
\omega_{d\sigma}&=-E_d-U\sum^{\sigma^\prime\not=\sigma}_{\sigma^\prime}\left(1-\langle n_{d\sigma^\prime}\rangle\right).
\end{align}
Thus, the single-hole energy of $d$ holes is defined as follows:
\begin{align}
E_{0S}&=\frac{E_0}{n_{td}},
\end{align}
where $E_0=\sum_{\sigma}\omega_{d\sigma}\langle n_{d\sigma}\rangle$, and $n_{td}=\langle n_{d\frac{1}{2}}\rangle+\langle n_{d-\frac{1}{2}}\rangle+\langle n_{d\frac{3}{2}}\rangle+\langle n_{d-\frac{3}{2}}\rangle$.

\begin{figure}
    \centering
    \includegraphics[width=0.4\textwidth]{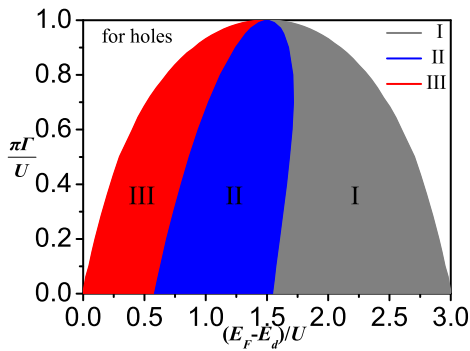}
    \caption{(Color online) Mean-field magnetic phase diagram for holes.}
\end{figure}

The calculation of the LDOS and ground state energy for holes is similar to that for particles, the corresponding analysis is not repeated for the sake of brevity. Here, the mean-field magnetic phase diagram for holes is presented in Fig. 13. Similar to Fig. 8, Fig. 13 shows that the magnetic phases III, II, and I correspond to three, two, and one hole occupation, respectively. Furthermore, the results in Fig. 13 and those in Fig. 8 are symmetrical with $(E_F-E_d)/U=1.5$ as a line axis; the region I, II, and III for holes are interpreted as the region I, II, and III for particles, signalling the PH symmetry of the model.

\end{appendices}


\begin{thebibliography}{}
\bibitem{1} P. W. Anderson, Phys. Rev. {\bf 124}, 41 (1961).
\bibitem{2} A. Georges, C. R. Physique {\bf 17}, 430 (2016).
\bibitem{3} J. Friedel, Nuovo Cimento {\bf 7}, 287 (1958).
\bibitem{4} J. Kondo, Prog. Theor. Phys. {\bf 32}, 37 (1964).
\bibitem{5} H. G. Luo, T. Xiang, X. Q. Wang, Z. B. Su, and L. Yu, Phys. Rev. Lett. {\bf 92}, 256602 (2004).
\bibitem{6} J. R. Schrieffer and P. A. Wolff, Phys. Rev. {\bf 149}, 491 (1966).
\bibitem{7} J. A. Appelbaum and D. R. Penn, Phys. Rev. {\bf 188}, 874 (1969).
\bibitem{8} A. Theumann, Phys. Rev. {\bf 178}, 978 (1969).
\bibitem{9} K. G. Wilson, Rev. Mod. Phys. {\bf 47}, 773 (1975).
\bibitem{10} C. Lacroix, J. Phys. F: Metal Phys. {\bf 11}, 2389 (1981).
\bibitem{11} C. Lacroix, J. Appl. Phys. {\bf 53}, 2131 (1982).
\bibitem{12}  P. Schlottmann, Z. Phy. B-Condensed Matter. {\bf 52}, 127 (1983).
\bibitem{13} P. Schlottmann, Phys. Rev. Lett. {\bf 50} 1697 (1983).
\bibitem{14} J. A. White, Phys. Rev. B {\bf 45}, 1100 (1992).
\bibitem{15} P. Gambardella, S. S. Dhesi, S. Gardonio, C. Grazioli, P. Ohresser, and C. Carbone, Phys. Rev. Lett. {\bf 88}, 047202 (2002).
\bibitem{16} M. Nuss, E. Arrigoni, M. Aichhorn, and W. von der Linden, Phys. Rev. B {\bf 85}, 235107 (2012).
\bibitem{17} M. Sch\"{u}ler, C. Renk, and T. O. Wehling, Phys. Rev. B {\bf 91}, 235142 (2015).
\bibitem{18} N. S. Maslova, P. I. Arseyev, and V. N. Mantsevich, Solid State Commun. {\bf 252}, 78 (2017).
\bibitem{19} S. Mukherjee and D. R. Reichman, Phys. Rev. B {\bf 95}, 155111 (2017).
\bibitem{20} L. Joly, J.-P. Kappler, P. Ohresser, Ph. Sainctavit, Y. Henry, F. Gautier, G. Schmerber, D. J. Kim, C. Goyhenex, H. Bulou, O. Bengone, J. Kavich, P. Gambardella, and F. Scheurer, Phys. Rev. B {\bf 95}, 041108(R) (2017).
\bibitem{21} K. Hirsch, V. Zamudio-Bayer, A. Langenberg, M. Niemeyer, B. Langbehn, T. M\"{o}ller, A. Terasaki, B. v. Issendorff, and J. T. Lau, Phys. Rev. Lett. {\bf 114}, 087202 (2015).
\bibitem{22} K. Hirsch, J. T. Lau, and B. v. Issendorff, arXiv: 1407.2018v1, 2014.
\bibitem{23} B. Uchoa, V. N. Kotov, N. M. R. Peres, and A. H. Castro Neto, Phys. Rev. Lett. {\bf 101}, 026805 (2008).
\bibitem{24} M. Mashkoori, I. Mahyaeh, and S. A. Jafari, J. Phys. Soc. Jpn. {\bf 85}, 014707 (2016).
\bibitem{25} A. S. Rodin and A. H. Castro Neto, Phys. Rev. B {\bf 97}, 235428 (2018).
\bibitem{26} L. Riegger, N. Darkwah Oppong, M. H\"{o}fer, D. R. Fernandes, I. Bloch, and S. F\"{o}lling, Phys. Rev. Lett. {\bf 120}, 143601 (2018).
\bibitem{27} I. Kuzmenko, T. Kuzmenko, Y. Avishai, and G. B. Jo, Phys. Rev. B {\bf 97}, 075124 (2018).
\bibitem{28} C. Wu, J.-P. Hu, and S.-C. Zhang, Phys. Rev. Lett. {\bf 91}, 186402 (2003).
\bibitem{29} C. Wu, Phys. Rev. Lett. {\bf 95}, 266404 (2005).
\bibitem{30} Y. Dong and H. Pu, Phys. Rev. A {\bf 87}, 043610 (2013).
\bibitem{31} J. S. Krauser, U. Ebling, N. Fl\"{a}schner, J. Heinze, K. Sengstock, M. Lewenstein, A. Eckardt, and C. Becker, Science {\bf 343}, 157 (2014).
\bibitem{32} U. Ebling, J. S. Krauser, N. Fl\"{a}schner, K. Sengstock, C. Becker, M. Lewenstein, and A. Eckardt, Phys. Rev. X {\bf 4}, 021011 (2014).
\bibitem{33} T.-L. Ho and B. Huang, Phys. Rev. A {\bf 91}, 043601 (2015).
\bibitem{34} Z. Sun and Q. Gu, Sci. Rep. {\bf 6}, 31776 (2016).
\bibitem{35} J. Xu, T. Feng, and Q. Gu, Annals of Physics {\bf 379}, 175 (2017).
\bibitem{36} E. Szirmai, G. Barcza, J. S\'{o}lyom, and \"{O}. Legeza, Phys. Rev. A {\bf 95}, 013610 (2017).
\bibitem{37} M. R. Galpin, A. B. Gilbert, and D. E. Logan, J. Phys.: Condens. Matter {\bf 21}, 375602 (2009).
\bibitem{38} M. R. Galpin and D. E. Logan, Phys. Rev. B {\bf 77}, 195108 (2008).
\bibitem{39} S. Sch\"{a}fer, Phys. Rev. B {\bf 83}, 195110 (2011).
\end{thebibliography}
\end{document}